\documentclass[aps,prd,superscriptaddress]{revtex4}
\usepackage{epsfig,epsf}
\usepackage{amsmath}
\usepackage{amsthm}
\usepackage{amsfonts}
\usepackage{amssymb}
\usepackage{dsfont}
\usepackage{multirow}
\usepackage{appendix}
\usepackage{slashed}
\usepackage[active]{srcltx}
\usepackage{psfrag}
\usepackage{subfigure}
\begin{document}
\title{\Large{\bf { Analysis of $D_s D K^*$ and $D_s D^{*} K^*$  vertices  and branching ratio of $B^+\to
{K^*}^0 \pi^+$ }}}

\author{\small M. Janbazi$^1$ \footnote {e-mail: mehdijanbazi@yahoo.com}
R. Khosravi$^2$  \footnote {e-mail: rezakhosravi@cc.iut.ac.ir} }

\affiliation{\emph{$^1$Department of
Physics, Shiraz Branch Islamic Azad University, Shiraz, Iran}\\\emph{$^2$Department of Physics, Isfahan University of
Technology, Isfahan 84156-83111, Iran } }

\begin{abstract}
In this paper, the strong form factors and coupling constants of
$D_sDK^*$ and $D_sD^*K^*$  vertices are investigated within the
three-point QCD sum rules method with and without the $SU_{f}(3)$
symmetry. In this calculation, the contributions of the quark-quark,
quark-gluon, and gluon-gluon condensate corrections are considered.
As an example of specific application of these coupling constants,
the branching ratio of the hadronic decay $B^+\to {K^*}^0 \pi^+$ is
analyzed based on the one-particle-exchange which is one of the
phenomenological models. In this model, $B$ decays into a $D_s D^*$
intermediate state, and then these two particles exchange a $D
(D^*)$ producing the final $K^*$ and $\pi$ mesons. In order to
compute the effect of these interactions, the $D_s D K^*$ and $D_s
D^* K^*$ form factors are needed.
\end{abstract}


\maketitle

\section{Introduction}
In high energy physics, investigation of meson interactions depends
on information about the proper functional form of strong form
factors. Among all vertices, the charmed meson ones, which play an
important role in understanding the final-state re-scattering
effects in the hadronic $B$ decays, are much more significant. They
are related to the basic parameters $\beta$ and $\lambda$ in the
heavy quark effective Lagrangian \cite{HQEFT}. Therefore,
researchers have concentrated on computing the strong form factors
and coupling constants connected to these vertices. Until now, the
vertices involving charmed mesons such as $D^* D^* \rho$
\cite{MEBracco}, $D^* D \pi$ \cite{FSNavarra,MNielsen}, $D D \rho$
\cite{MChiapparini}, $D^* D \rho$ \cite{Rodrigues3}, $D D J/\psi$
\cite{RDMatheus},  $D^* D J/\psi$ \cite{RRdaSilva}, $D^*D_sK$,
$D^*_sD K$, $D^*_0 D_s K$, $D^*_{s0} D K$ \cite{SLWang}, $D^*D^* P$,
$D^*D V$, $D D V$ \cite{ZGWang}, $D^* D^* \pi$ \cite{FCarvalho},
$D_s D^* K$, $D_s^* D K$ \cite{ALozea}, $D D \omega$
\cite{LBHolanda},  $D_s D_s V$, $D^{*}_s D^{*}_s V$ \cite{KJ,KJ2},
and $D_1D^*\pi, D_1D_0\pi, D_1D_1\pi$ \cite{Janbazi} have been
studied within the framework of the QCD sum rules.

The effective Lagrangians for the interaction $D_sDK^*$ and
$D_sD^*K^*$  vertices are as follows: \cite{BraChia}:
\begin{eqnarray}\label{eq11}
{\cal L}_{D_s D K^*}&=&i g_{D_s D K^*}~{K^*}^\alpha (\bar D_s
\partial_\alpha D  -\partial_\alpha \bar D_s D),\nonumber \\
{\cal L}_{D_s D^* K^*}&=&-g_{D_s D^* K^*} \epsilon^{ \alpha\beta
\rho \sigma}\partial_{\alpha} D^*_{\beta}(\partial_{\rho}
K^{*}_{\sigma} \bar D_s +D_s \partial_{\rho} \bar K^{*}_{\sigma}),
\end{eqnarray}
where $g_{D_s D K^*}$ and $g_{D_s D^* K^*}$ are  the strong form
factors. From these Lagrangians, the elements related to the
$D_sDK^*$ and $D_sD^*K^*$ vertices can be derived in terms of the
strong form factors as:
\begin{eqnarray}\label{eq12}
\langle D(p) D_s(p')   |K^*(q, \varepsilon'')   \rangle &=& -g_{D_s
D K^*}(q^2)\times(p^\mu+p'^\mu)\varepsilon''_\mu ,\nonumber\\
\langle D^*(p,\varepsilon) D_s(p') |   K^*(q,\varepsilon'') \rangle &=& i
g_{D_s D^* K^*}(q^2)\times \epsilon^{ \alpha\beta \mu \nu}
p'_\alpha q_\beta \varepsilon_\mu(p) \varepsilon''_\nu(q) ,
\end{eqnarray}
where $q=p-p'$.

In this work, we decide to calculate the strong form factors and
coupling constants associated with the $D_sDK^*$ and $D_sD^*K^*$
vertices in the frame work of the three-point QCD sum rules (3PSR).
As an example of specific application of these coupling constants
can be pointed out to branching ratio calculations of hadronic $B$
decays. I this paper, we would like to consider the branching ratio
of the $B^+\to {K^*}^0 \pi^+$ decay according to the coupling
constants of the $D_sDK^*$ and $D_sD^*K^*$ vertices.

The plan of the present paper is as follows: In Section II, the
strong form factor calculation of the $D_{s} D K^*$ vertex is
derived in the framework of the 3PSR;   computing the quark-quark,
quark-gluon and gluon-gluon condensate contributions in the Borel
transform scheme. Using necessary changes in the expression obtained
for the $g_{D_{s} D K^*}$, the strong form factor $g_{ D_{s} D^{*}
K^*}$ is presented. In Section III, we analyze the strong form
factors as well as the coupling constants with and without the
$SU_{f}(3)$ symmetry. For a better analysis, a comparison is made
between  our results and the predictions of other methods. Finally,
we consider the branching ratio of the $B^+\to {K^*}^0 \pi^+$ decay
using the coupling constants of the $D_sDK^*$ and $D_sD^*K^*$
vertices.

\section{STRONG FORM FACTORS OF $D_sDK^*$ and $D_sD^*K^*$  vertices}
\begin{figure}[th]
\includegraphics[width=8cm,height=2.5cm]{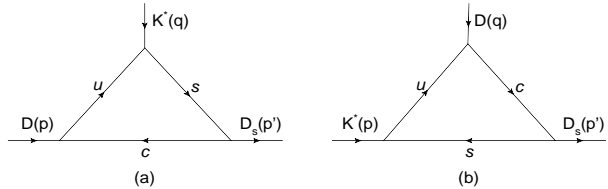}
\caption{Perturbative diagrams for off-shell $K^*$ (a) and off-shell
$D$ meson (b).}\label{F1}
\end{figure}

To compute the strong form factors of the $D_sDK^*$ and $D_sD^*K^*$
vertices via the 3PSR, we start with the following correlation
functions as
\begin{eqnarray}\label{eq21}
\Pi^{K^*}_{\mu}(p, p')&=& i^2 \int d^4x  d^4y e^{i(p'x-py)}\langle 0
|\mathcal{T}\left\{j^{D_s}(x) {j_\mu^{K^*}}^{\dagger}(0)
{j^{D}}^{\dagger}(y)\right\}| 0 \rangle,\nonumber \\
\Pi^{K^*}_{\mu\nu}(p, p')&=& i^2 \int d^4x  d^4y
e^{i(p'x-py)}\langle 0 |\mathcal{T}\left\{j^{D_s}(x)
{j_\mu^{K^*}}^{\dagger}(0) {j^{D^*}_{\nu}}^{\dagger}(y)\right\}| 0
\rangle,
\end{eqnarray}
where $K^*$ is supposed as an off-shell meson. For off-shell charmed
mesons, the correlation functions are:
\begin{eqnarray}\label{eq22}
\Pi^{D}_{\mu}(p, p') &=& i^2 \int d^4x  d^4y e^{i(p'x-py)}\langle 0
|\mathcal{T}\left\{ j^{D_s}(x){j^{D}}^{\dagger}(0){j_{\mu}^{K^*}}^{\dagger}(y)
\right\}| 0 \rangle,\nonumber \\
\Pi^{D^*}_{\mu\nu}(p, p') &=& i^2 \int d^4x  d^4y e^{i(p'x-py)}\langle 0
|\mathcal{T}\left\{ j^{D_s}(x){j^{D^*}_{\nu}}^{\dagger}(0){j_{\mu}^{K^*}}^{\dagger}(y)
\right\}| 0 \rangle,
\end{eqnarray}
where $j^{D_{s}}=\bar c \gamma_5 s,~j^{D}=\bar c \gamma_5 u,~j^{D^*_{\mu}}=\bar c \gamma_{\mu} u$, and
$j^{K^*}_{\mu}=\bar u \gamma_{\mu} s $ are interpolating currents
with the same quantum numbers of $D_{s},~D,~D^*$, and $K^*$ mesons,
respectively. Also, $\mathcal{T}$ is time ordering product, $p$ and
$p'$ are the four momentum of the initial and final mesons,
respectively, as depicted in Fig. \ref{F1}.

To calculate the strong form factor of the $D_{s} D K^*$ vertex in
the framework of the 3PSR, the correlation functions in Eqs.
(\ref{eq21}) and (\ref{eq22}) are calculated in two different ways.
First, they are calculated in the space-like region in terms of
quark-gluon language like quark-quark, gluon-gluon condensate, etc.
using the Wilson operator product expansion (OPE). It is called the
QCD or theoretical side of  the QCD sum rules. Second, in the
hadronic representation, they are calculated in the time-like region
in terms of hadronic parameters such as the form factors, decay
constants and masses. It is named the phenomenological or physical
side.

In order to calculate the phenomenological part of the correlation
functions in Eqs. (\ref{eq21}) and (\ref{eq22}), three complete sets
of intermediate states with the same quantum number should be
inserted in these equations. Performing the Fourier transformation,
for the phenomenological parts, we have:
\begin{eqnarray}\label{eq28}
\Pi^{K^*}_{\mu}&=&\frac{\langle 0|j^{D_s}|D_s(p')\rangle\langle
0|j^{D }|D(p)\rangle\langle D_s(p')D(p)|K^*(q,\epsilon)\rangle
\langle K^*
(q,\epsilon)|j^{K^*}_{\mu}|0\rangle}{(p^2-m^2_{D})(p'^2-m^2_{D_s})(q^2-m^2_{K^*})}
\nonumber\\&&+\mbox{higher and continuum states,}\nonumber\\
\Pi^{D}_{\mu}&=&\frac{\langle 0|j^{D_s}|D_s(p')\rangle\langle
0|j_\mu^{K^*} |K^*(p,\epsilon)\rangle\langle
 D_s(p')K^*(p,\epsilon)|D(q) \rangle \langle D
(q)|j^{D}|0\rangle}{(p^2-m^2_{K^*})(p'^2-m^2_{D_s})(q^2-m^2_{D})}
\nonumber\\&&+\mbox{higher and continuum states.}
\end{eqnarray}
The matrix elements $\langle 0 | j_{\mu}^{K^*} | K^*(q,\varepsilon)
\rangle$, and  $\langle 0 | j^{D_{(s)}} | D_{(s)}(p) \rangle$  are
defined as:
\begin{eqnarray}\label{eq29}
\langle 0 | j_{\mu}^{K^*} | K^*(q,\varepsilon) \rangle &=& m_{K^*}
f_{K^*} \varepsilon_{\mu}(q), \nonumber \\
\langle 0 | j^{D_{(s)}} | D_{(s)}(p) \rangle &=& \frac{
m^2_{D_{(s)}} f_{D_{(s)}}}{m_c+m_{u(s)}},
\end{eqnarray}
where $m_{K^*}$, $m_{D_{(s)}}$, $f_{K^*}$,  and  $f_{D_{(s)}}$  are
the masses and decay constants of mesons $K^*$  and $D_{(s)}$,
respectively. $\varepsilon_\mu$ is the polarization vector of the
vector  meson $K^*$.

Inserting Eqs. (\ref{eq12}) and (\ref{eq29}) in Eq. (\ref{eq28}) and
after some calculations, we obtain $\Pi_{\mu}^{K^*}$ and
$\Pi_{\mu}^{D} $ in terms of the strong form factors $g^{K^*}_{D_s D
K^* }$ and $g^{D}_{D_s D K^* }$ as:
\begin{eqnarray}\label{eq210}
\Pi_{\mu}^{K^*} &=& -g^{K^*}_{D_s D K^* }(q^2)\frac{ m_{K^*}m_{D}^2
m_{D_s}^2 f_{K^*}f_{D} f_{D_s}
}{m_c(m_c+m_s)(p^2-m^2_{D})(p'^2-m^2_{D_s})(q^2-m^2_{K^*})}p_{\mu}\nonumber\\&&+\mbox{higher and continuum states,}\,\nonumber\\
\Pi_{\mu}^{D}  &=&-g^{D}_{ D_s D K^* }(q^2)\frac{m_{D_s}^2m_D^2
f_{K^*}f_{D}f_{D_s}(m_{D}^2+m_{K^*}^2-q^2)}{m_c(m_c+m_s)m_{K^*}
(p^2-m_{K^*}^2)(p'^2-m_{D_s}^2)(q^2-m_{D}^2)}p_{\mu}\nonumber\\&&+\mbox{higher
and continuum states}\,.
\end{eqnarray}

In the theoretical side, the three-point correlation function
contains the perturbative and nonperturbative parts as
\begin{eqnarray}\label{eq27}
\Pi^{K^*(D)}_{\mu}=(\Pi^{K^*(D)}_{per}+\Pi^{K^*(D)}_{nonper}) ~p_\mu
+\mbox{other structures\,.}
\end{eqnarray}
According to the 3PSR method, we can estimate the perturbative part
of the correlation function, using  the double dispersion relation,
as
\begin{eqnarray}\label{eq24}
\Pi_{per}^{K^*(D)}= -\frac{1}{4 \pi^2} \int ds\int
ds'\frac{\rho^{K^*(D)}}{(s-p^2)(s'-p'^2)}+\mbox{subtraction
terms}\,,
\end{eqnarray}
where $\rho^{K^*(D)}$ is spectral density. The spectral density is
calculated in terms of the usual Feynman integrals by the help of
the Cutkosky rules, where the quark propagators are replaced by
Dirac-delta functions, i.e., $\frac{1}{p^2-m^2}\rightarrow(-2\pi
i)\delta(p^2-m^2)$. The diagrams corresponding to the perturbative
part (bare loop) are depicted in Fig. \ref{F1}.
Using Fig. \ref{F1} and after some straightforward calculations, we
have:

$\bullet$ For the off-shell $K^*$ (Fig. \ref{F1} (a)):
\begin{eqnarray*}\label{eq25}
\rho^{K^*}_{ D_s D K^* }&=&6I_0[2 m_cm_s-2m_c^2+\Delta'+ C'_1 (2
m_cm_s-2m_c^2+u) ].
\end{eqnarray*}

$\bullet$ For the off-shell $D$ (Fig. \ref{F1} (b)):
\begin{eqnarray*}\label{eq26}
\rho^{D}_{ D_s D K^*}&=&6I_0[2m_cm_s-2m_s^2+\Delta+ C_1
(2m_cm_s-2m_s^2+2\Delta+ u) ].
\end{eqnarray*}
The explicit expressions of the coefficients in the spectral
densities  are given in Appendix-A.

Now, the non-perturbative part contributions to the correlation
function are discussed. In QCD, the correlation function can be
evaluated by OPE in the deep Euclidean region. Using the expansion
of it in terms of a series of local operators with increasing
dimension, we get:
\begin{eqnarray}\label{eq23}
\Pi_{\mu}&=& C_{\mu}^{(0)}{\rm
I}+C_{\mu}^{(3)}\langle0|\bar{\Psi}\Psi|0\rangle+C_{\mu}^{(4)}\langle0|G^{a}_{\rho\nu}G_{a}^{\rho\nu}|0\rangle+C_{\mu}^{(5)}\langle
0| \bar{\Psi} \sigma_{\rho\nu} T^a G_a^{\rho\nu} \Psi| 0\rangle
\nonumber\\
&+&
C_{\mu}^{(6)}\langle0|\bar{\Psi}\Gamma\Psi\bar{\Psi}\Gamma'\Psi|0\rangle+...,
\end{eqnarray}
where $C_{\mu}^{(i)}$ are the Wilson coefficients, I is the unit
operator, $\bar{\Psi}$ is the local fermion field operator and
$G^{\rho\nu}$ is the gluon strength tensor. The Wilson coefficient
$C_{\mu}^{(0)}$ is the contribution of the perturbative part of QCD
( i.e., $\Pi^{K^*(D)}_{per}$), and the other coefficients are
contributions of the non-perturbative part.

In Eq. (\ref{eq23}), the Wilson coefficients $C_{\mu}^{(3)}$,
$C_{\mu}^{(4)}$ and $C_{\mu}^{(5)}$ of dimensions $3$, $4$ and $5$
are related to contributions of the quark-quark, gluon-gluon and
quark-gluon condensate, respectively. Also, $C_{\mu}^{(6)}$ is
connected to contribution of the four-quark condensate of dimension
six. For the calculation of the condensate terms, we consider these
points:

a) Our calculations show that the contributions of the four-quark
condensate are less than a few percent, therefore the condensate
terms of dimensions $3, 4$ and $ 5$ are more important than the
other terms in OPE.

b) In the 3PSR, when the light quark  is a spectator, the
gluon-gluon condensate contributions can be easily ignored
\cite{Colangelo}.

c) The quark condensate contribution of the light quark which is a
non spectator, is zero after applying the double Borel
transformation with respect to the both variables $p^2$ and
$p'^{2}$, because only one variable appears in the denominator.

d) In the 3PSR, when the heavy quark  is a spectator, the
quark-quark condensate contributions are suppressed by inverse of
the heavy quark mass, and can be safely omitted \cite{Colangelo}.

\begin{figure}[th]
\includegraphics[width=8cm,height=2cm]{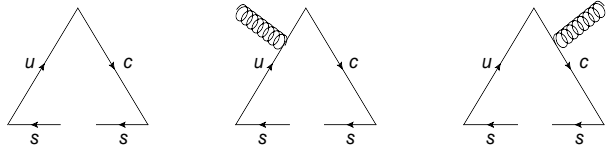}
\caption{Non-perturbative diagrams for the off-shell $D$
meson.}\label{F2}
\end{figure}

Therefore, to compute the contribution of the non-perturbative part
of the correlation function for the off-shell $D$ meson, three
diagrams of dimensions $3$ and $5$, shown in Fig. \ref{F2}, are
considered. In this case, the quark-quark and quark-gluon diagrams
are more important than the other terms in the OPE since the light
quark $s$  is a spectator.

\begin{figure}[th]
\includegraphics[width=7cm,height=6cm]{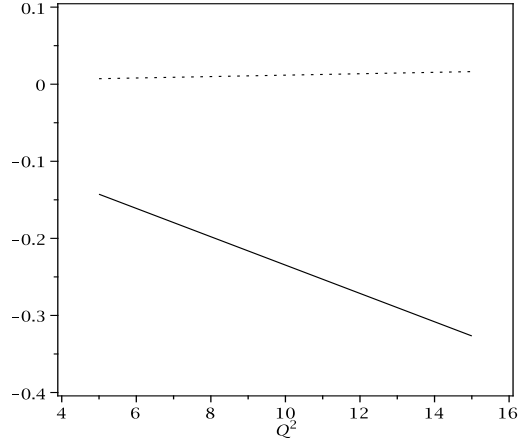}
\caption{The quark-quark  plus quark-gluon condensate contributions
(solid line) and also gluon-gluon condensate (dot line) on $Q^2$ for
the off-shell $D$ meson.} \label{F3}
\end{figure}

For a better analysis, we compare the contributions of the
quark-quark plus the quark-gluon condensates with the gluon-gluon
condensate for the off-shell $D$ meson in Fig. \ref{F3}, in the
interval $5~\rm{GeV}^2 \leq Q^2\leq 15~\rm{GeV}^2$~ $(Q^4=-q^2)$. As
can be seen, the gluon-gluon condensate contributions can be easily
ignored.

\begin{figure}[th]
\includegraphics[width=7cm,height=4cm]{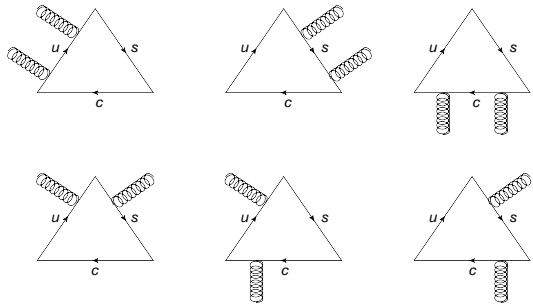}
\caption{Non-perturbative diagrams for the off-shell $K^*$
meson.}\label{F4}
\end{figure}

When $K^*$ is an off-shell meson, the gluon-gluon diagrams of
dimension $4$ are more important than the quark-quark and
quark-gluon condensates since the heavy quark $c$  is a spectator.
Fig. \ref{F4} shows these diagrams related to the gluon-gluon
condensate.

We show our results for the non-perturbative contributions
corresponding to Figs. \ref{F2} and \ref{F4} as
\begin{eqnarray}\label{eq27}
C_{\mu}^{(4)}= \frac{C_{D_sDK^*}^{K^*}}{12}~ p_{\mu},~~~~~
C_{\mu}^{(3)}+C_{\mu}^{(5)}=\frac{C_{D_sDK^*}^{D}}{12}~ p_{\mu}.
\end{eqnarray}
where the explicit expressions for $C_{D_sDK^*}^{K^*}$ and
$C_{D_sDK^*}^{D}$  are given in Appendix-B. It should be noted that
in order to obtain the gluon-gluon condensate contributions, we will
follow the same procedure as stated in \cite{Kiselev}.

The strong form factors are calculated by equating two
representations of the correlation function and applying the Borel
transformations with respect to the $p^2(p^2\rightarrow M^2_1)$ and
$p'^2(p'^2\rightarrow M^2_2)$ on the phenomenological as well as the
perturbative and nonperturbative parts of the correlation function
in order to suppress the contributions of the higher states and
continuum. The equations for the strong form factors are obtained as
follows:
\begin{eqnarray}\label{eq212}
g^{K^*}_{D_sDK^* }(q^2)&=&\Lambda^{K^*}_{D_sDK^*}
\left\{-\frac{1}{4\pi^2}\int^{s_0^{D_s}}_{(m_c+m_s)^2}ds'
\int^{s_0^D}_{s_1} ds \rho^{K^*}_{D_sDK^*}(s,s',q^2)
e^{-\frac{s}{M_1^2}}
e^{-\frac{s'}{M_2^2}} \right.\nonumber \\
&-&\left. iM^{2}_{1}M^{2}_{2} \left \langle \frac{\alpha_s}{\pi} G^2
\right\rangle \times C^{(4)} \right\},\nonumber\\
g^{D}_{D_sDK^*}(q^2)&=&  \Lambda^{D}_{D_sDK^*}
\left\{-\frac{1}{4\pi^2}\int^{s^{D_s}_0}_{(m_c +m_s)^2}ds'
\int^{s^{K^*}_0}_{s_2} ds \rho^{D}_{ D_sDK^*}(s,s',q^2)
e^{-\frac{s}{M_1^2}} e^{-\frac{s'}{M_2^2}} \right. \nonumber \\
&+& \left.  M_1^2 M_2^2~  \langle s\bar s\rangle
\times(C^{(3)}+C^{(5)}) \right\},
\end{eqnarray}
where $s^{K^*}_0$ and $s^{D(D_s)}_0$ are the continuum thresholds in
$K^*$ and $D(D_s)$ mesons, respectively. $s_1$ and $s_2$ are the
lower limits of the integrals over $s$ as
\begin{eqnarray*}\label{eq225}
s_{1}=\frac{m_c^2(m_c^2-s'+q^2)}{m_{c}^{2}-s'}\,,\quad\quad\quad
s_{2}=\frac{m_s^2(m_s^2-s'+q^2)}{m_{s}^{2}-s'}\,.
\end{eqnarray*}
Also $\Lambda^{K^*}_{D_sDK^*}$ and $\Lambda^{D}_{D_sDK^*}$  are
defined as:
\begin{eqnarray}\label{eq213}
\Lambda^{K^*}_{D_sDK^*}&=&-\frac{m_c(m_c+m_s)(q^2-m_{K^*}^{2})}{m_{K^*}
m_{D}^2m_{D_s}^2
f_{K^*}f_{D}f_{D_s}}~ e^{\frac{m_{D}^2}{M_1^2}}e^{\frac{m_{D_s}^2}{M_2^2}},\nonumber\\
\Lambda^{D}_{D_sDK^*} &=&-\frac{m_c(m_c+m_s)m_{K^*}(q^2-m_{D}^{2})}{
m_{D}^2m_{D_s}^2 f_{K^*}f_{D} f_{D_s}(m_{D_s}^2+m_{K^*}^2-q^2)}
~e^{\frac{m_{K^*}^2}{M_1^2}} e^{\frac{m_{D_s}^2}{M_2^2}}.
\end{eqnarray}

Following  the previous steps, relations similar  to Eq.
(\ref{eq212}) can be obtained for the strong form factors of the
$D_{s} D^{*} K^*$ vertex via the 3PSR. It should be noted that in
this case, calculations are done for the Lorentz structure
$\epsilon^{ \alpha\beta \mu \nu}p_{\alpha}p'_{\beta}$. In order to
have the correct relations for the strong form factors of the $D_{s}
D^{*} K^*$ vertex, the appropriate terms of $\Lambda$, the spectral
density $\rho$, and quark-gluon condensate $ C^{(4)}=
\frac{C_{D_sD^*K^*}^{K^*}}{12}$ and
$C^{(3)}+C^{(5)}=\frac{C_{D_sD^*K^*}^{D^*}}{12}$ should be replaced
in Eq. (\ref{eq212}). The explicit expressions for
$C_{D_sD^*K^*}^{K^*}$ and $C_{D_sD^*K^*}^{D^*}$  are given in
Appendix-B. In addition, proper expressions for $\Lambda$ related to
the strong form factors $g^{K^*}_{D_{s} D^{*} K^*}$ and
$g^{D^*}_{D_{s} D^{*} K^*}$ are as follows:
\begin{eqnarray}\label{eq216}
\Lambda^{K^*}_{D_{s}D^{*}K^*}&=&-\frac{(m_c+m_s)(q^2-m^2_{K^*})}{m_{K^*}m_{D^{*}}m^{2}_{D_{s}}
f_{K^*}f_{D^{*}}f_{D_{s}}}~ e^{\frac{m_{D^*}^2}{M_1^2}}e^{\frac{m_{D_s}^2}{M_2^2}},\nonumber \\
\Lambda^{D^{*}}_{D_{s}D^{*}K^*}&=&-\frac{(m_c+m_s)(q^2-m^2_{D^*})}{m_{K^*}m_{D^{*}}m^{2}_{D_{s}}
f_{K^*}f_{D^{*}}f_{D_{s}}}~e^{\frac{m_{K^*}^2}{M_1^2}}
e^{\frac{m_{D_s}^2}{M_2^2}}.
\end{eqnarray}
Also, the spectral densities are calculated as
\begin{eqnarray}\label{eq214}
\rho^{K^*}_{ D_s D^* K^* }&=&-12I_0[C'_1m_c+C'_2(m_c-m_s)+m_c],\nonumber \\
\rho^{D^*}_{ D_s D^* K^*}&=&12I_0[C_1m_s+C_2(m_s-m_c)+m_s].
\end{eqnarray}

\section{NUMERICAL ANALYSIS}
In this section, the strong form factors and coupling constants for
the $D_{s} D K^*$ and $D_s D^* K^*$ vertices as well as the
branching ratio  of  the $B^+\to {K^*}^0 \pi^+$ decay are
considered. For this aim, the values of quark and meson masses are
chosen as: $m_s = 0.14 \pm 0.01~\rm GeV$, $m_c=1.26 \pm 0.02 ~\rm
GeV$, $m_{D^*}=2.01~ \rm GeV$, $m_{K^*}=0.89~ \rm GeV$,
$m_{D_{s}}=1.97~\rm GeV$, and $m_{D}=1.87~\rm GeV$ \cite{PDG2012}.
Moreover, the leptonic decay constants are presented in Table
\ref{T31}.
\begin{table}[h]
\caption{The leptonic decay constants in $\rm MeV$. }\label{T31}
\begin{ruledtabular}
\begin{tabular}{cccc}
$f_{K^*}$ \cite{PDG2012}& $f_{D}$ \cite{ZGWang}& $f_{D_{s}}$ \cite{Artuso}& $f_{D^{*}}$ \cite{GLWang} \\
\hline $220 \pm 5$& $223 \pm 17$ &  $294 \pm 27$&$ 340\pm 12$
\end{tabular}
\end{ruledtabular}
\end{table}

There are four auxiliary parameters containing the Borel mass
parameters $M_1$ and $M_2$, and continuum thresholds $s^{K^*}_{0}$, $s_{0}^{D(D_s)}$ and $s^{D^*}_{0}$ in Eq. (\ref{eq212}). The strong form factors
and coupling constants are the physical quantities, and should be
independent of them. However the continuum thresholds are not
completely arbitrary; these are related to the energy of the first
exited state. The values of the continuum thresholds are taken to be
$s^{K^*}_{0}=(m_{K^*}+\delta)^2$, $s_{0}^{D(D_s)}=(m_{D(D_s)}+\delta')^2$ and $s_{0}^{D^*}=(m_{D^*}+\delta')^2$ . We use $0.50 ~\rm GeV\leq
\delta \leq 0.90~\rm \rm GeV$ and $0.30 ~\rm GeV\leq \delta'
\leq0.70~\rm \rm GeV$ \cite{FSNavarra,MNielsen,MEBracco}.

\begin{figure}[th]
\includegraphics[width=8cm,height=4cm]{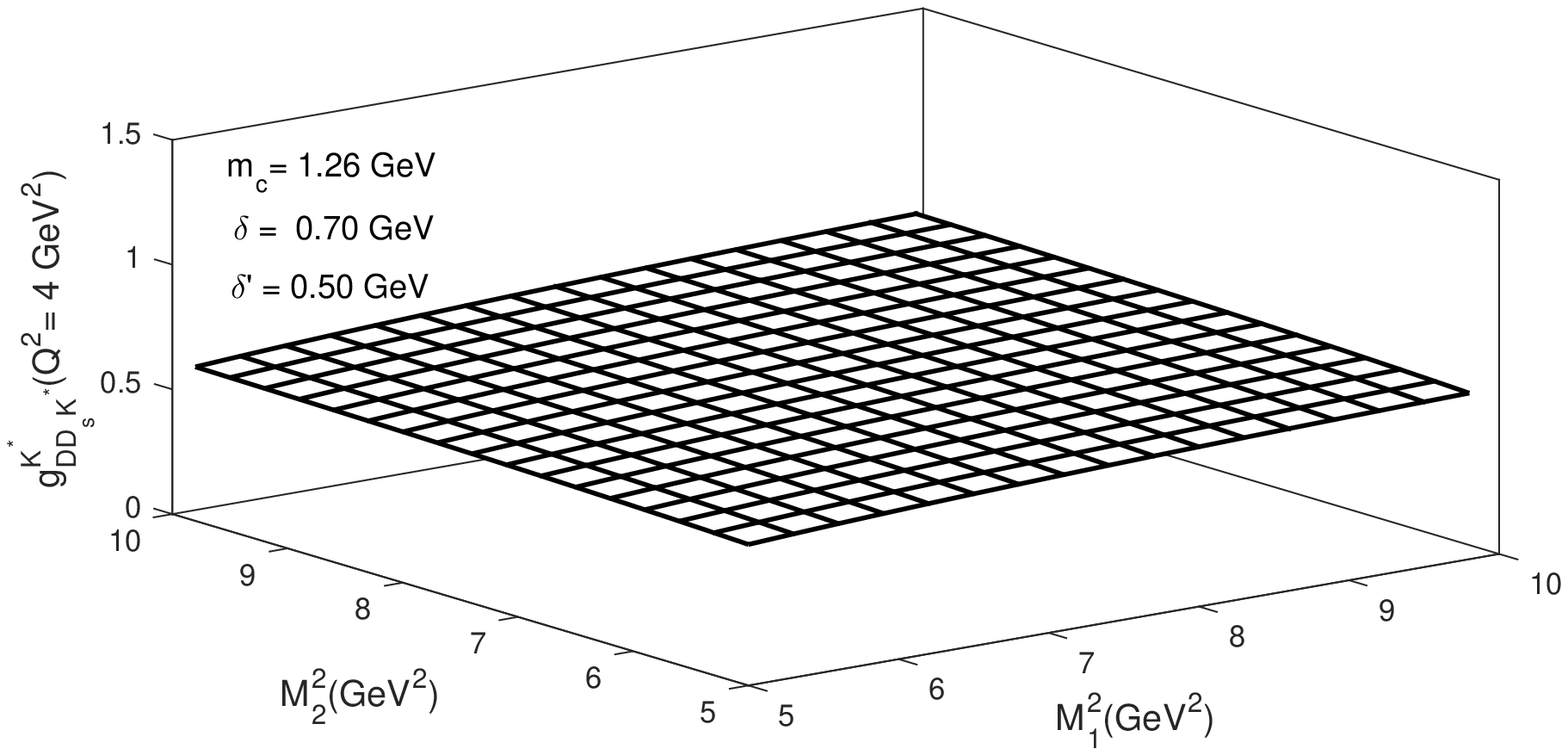}
\includegraphics[width=8cm,height=4cm]{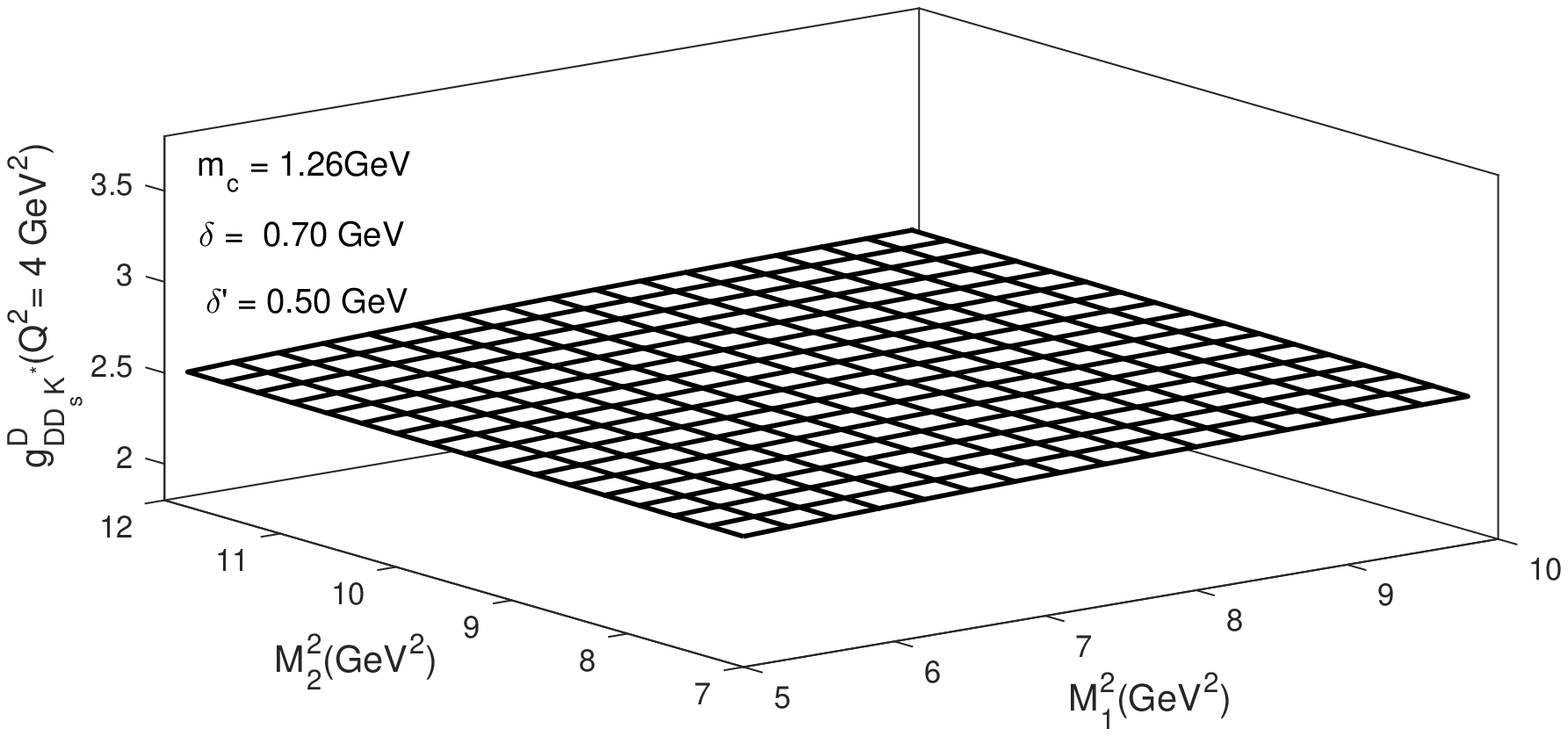}
\caption{$M_1^2$ and $M_2^2$ dependence of $g_{D_s D K^*}^{K^*}$ and
$g_{D_s D K^*}^{D}$.}\label{F301}
\end{figure}

\begin{figure}[th]
\includegraphics[width=8cm,height=4cm]{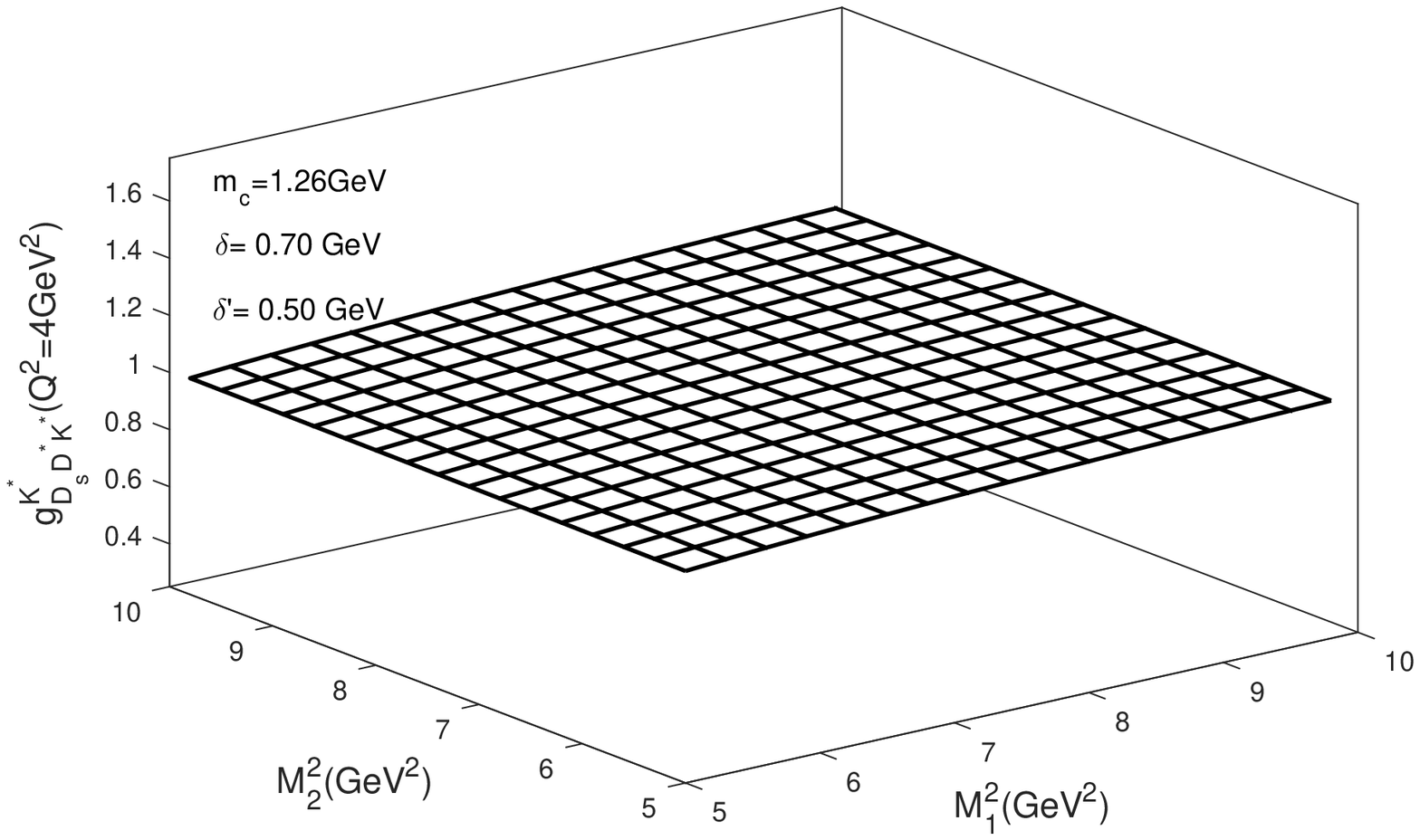}
\includegraphics[width=8cm,height=4cm]{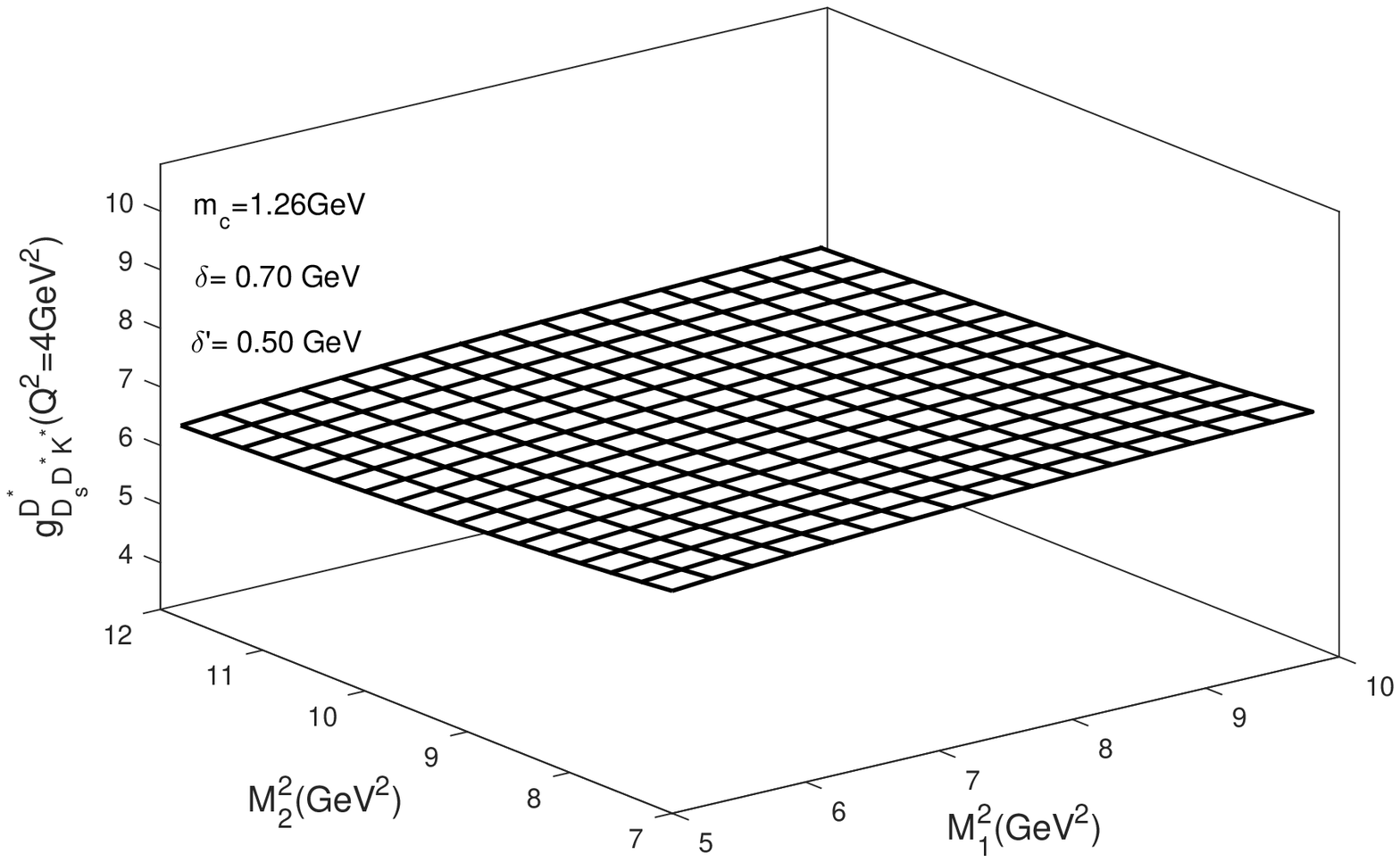}
\caption{$M_1^2$ and $M_2^2$ dependence of $g_{D_s D^* K^*}^{K^*}$
and $g_{D_s D^* K^*}^{D^*}$.}\label{F302}
\end{figure}

Our results should be almost insensitive to the intervals of the
Borel  parameters. On the other hand, the intervals of the Borel
mass parameters must suppress the higher states, continuum and
contributions of the highest-order operators. In other words, the
sum rule for the strong form factors must converge.  We get a very
good stability for the form factors as a function of the two
independent Borel parameters in the regions $5 ~{\rm GeV^2} < M_1^2
< 10 ~{\rm GeV^2} $ and $5 ~{\rm GeV^2} < M_2^2 < 10 ~{\rm GeV^2} $
when $K^*$ is an off-shell meson, and also $5 ~{\rm GeV^2} < M_1^2 <
10 ~{\rm GeV^2} $ and $7 ~{\rm GeV^2} < M_2^2 < 12 ~{\rm GeV^2} $
when $D$ ($D^*$) meson is an off-shell. For $Q^2=4~\rm{GeV}^2$, the
form factors $g_{D_s D K^*}^{K^*}$ and $g_{D_s D K^*}^{D}$, related
to the $D_sDK^*$ vertex, and $g_{D_s D^* K^*}^{K^*}$ and $g_{D_s D^*
K^*}^{D^*}$, connected to the $D_sD^*K^*$ vertex,  have been
illustrated with respect to the Borel parameters in Figs. \ref{F301}
and \ref{F302}, respectively. To calculate the strong form factors
$g_{D_s D K^*}^{K^*}$ and $g_{D_s D^* K^*}^{K^*}$, we get $[M_1^2,
M_2^2]=[7, 7]~\rm GeV^2$, and for $g_{D_s D K^*}^{D}$ and $g_{D_s
D^* K^*}^{D^*}$, we get $[M_1^2, M_2^2]=[7, 9]~\rm GeV^2$.

The numerical results for the strong form factors calculated via the
3PSR in Eq. (\ref{eq212}) have a cut-off. Therefore, we look for a
parametrization of the form factors in such that in the validity
region of the 3PSR, this parametrization coincides with the sum
rules prediction. Our numerical calculations show that the sum rule
predictions for the form factors in Eq. (\ref{eq212}) are well
fitted to the following function:
\begin{eqnarray*}\label{eq33}
g(Q^2)=A~e^{-Q^2/B}.
\end{eqnarray*}
The values of the parameters $A$ and $B$ are given in Table
\ref{T32} for various $(\delta, \delta')$.
\begin{table}
\caption{Parameters appearing in the fit functions for the $D_sDK^*$ and $D_sD^*K^*$
vertices for various $(\delta, \delta')$, where
$(\delta_1,\delta'_1)=(0.50,0.30),
~(\delta_2,\delta'_2)=(0.70,0.50)$ and
$(\delta_3,\delta'_3)=(0.90,0.70) ~\rm GeV$.}\label{T32}
\begin{ruledtabular}
\begin{tabular}{ccccccccc}
$\mbox{Form factor}$&$A(\delta_1,\delta'_1)$&$B(\delta_1,\delta'_1)$&$A(\delta_2,\delta'_2)$&$B(\delta_2,\delta'_2)$&$A(\delta_3,\delta'_3)$&$B(\delta_3,\delta'_3)$\\
\hline
$g^{K^*}_{D_sDK^*}(Q^2)$&1.90&2.28&2.42&2.97&2.93&4.01\\
$g^{D}_{D_sDK^*}(Q^2)$&2.36&31.02&3.02&30.68&3.28&30.87\\
$g^{K^*}_{D_sD^*K^*}(Q^2)$&3.76&11.40&3.97&6.79&4.09&3.55\\
$g^{D^*}_{D_sD^*K^*}(Q^2)$&3.65&54.77&4.18&42.05&4.74&28.42
\end{tabular}
\end{ruledtabular}
\end{table}

\begin{figure}[th]
\includegraphics[width=7cm,height=6cm]{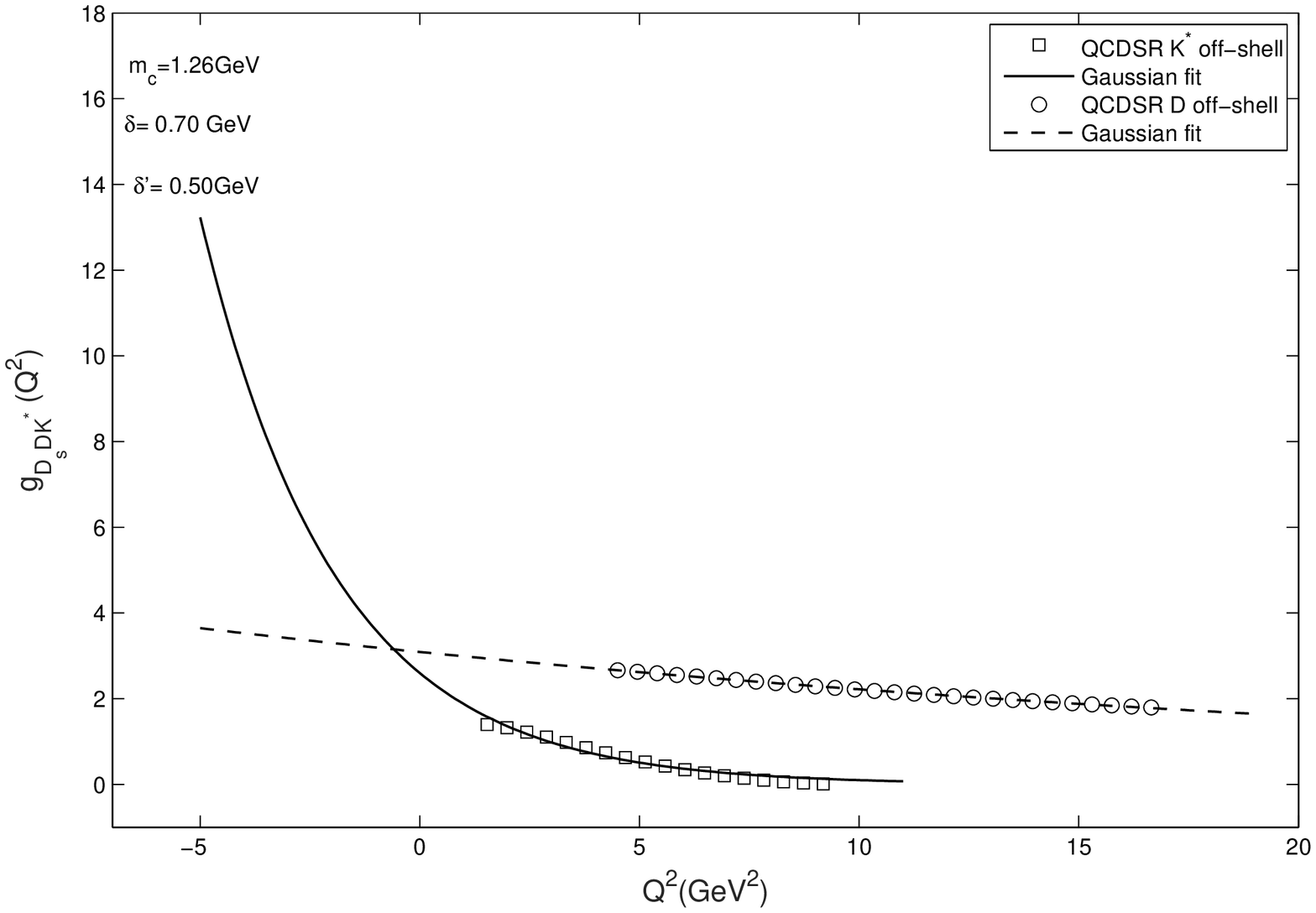}
\includegraphics[width=7cm,height=6cm]{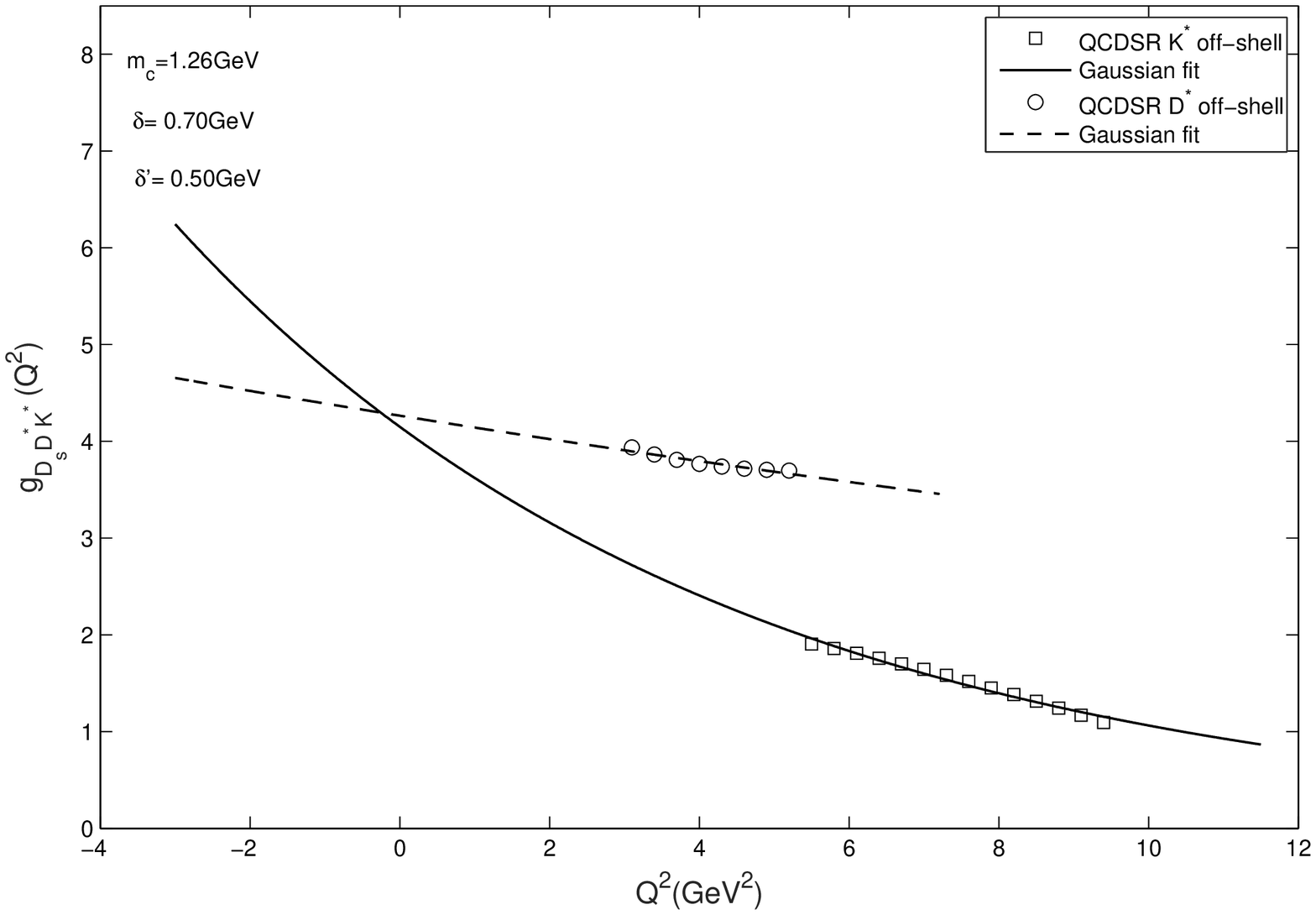}
\caption{The strong form factors $g^{D}_{D_sDK^*}$,
$g^{K^*}_{D_sDK^*}$, $g^{D^*}_{D_sD^*K^*}$ and $g^{K^*}_{D_sD^*K^*}$
on $Q^2$. }\label{F32}
\end{figure}

The dependence of the strong form factors $g^{D}_{D_sDK^*}(Q^2)$,
$g^{K^*}_{D_sDK^*}(Q^2)$, $g^{D^*}_{D_sD^*K^*}(Q^2)$ and
$g^{K^*}_{D_sD^*K^*}(Q^2)$  in $Q^2$ are shown in Fig. \ref{F32}.
The boxes and circles in Fig. \ref{F32} show the results of the
numerical evaluation via the 3PSR for the form factors $g^{K^*}_{
D_sDK^*} (g^{K^*}_{ D_sD^*K^*})$ and $g^{D}_{ D_sDK^*}(g^{D^*}_{
D_sD^*K^*})$, respectively. As can be seen, the form factors and
their fit functions coincide together, well.

The value of the strong form factors at $Q^2 = -m_m^2$, where $m_m$
is the mass of the off-shell meson, is defined as coupling constant.
Coupling constant results of the two vertices, ${D_sDK^*}$ and
${D_sD^*K^*}$, are presented in Table \ref{T33}. It should be
mentioned that the coupling constant  $g_{D_sDK^*}$ is the
dimensionless quantity and the coupling constant  $g_{D_sD^*K^*}$ is
in the unit of $\rm GeV^{-1}$.
\begin{table}[th]
\caption{The  coupling constant of the vertices ${D_sDK^*}$ and ${D_sD^*K^*}$,
}\label{T33}
\begin{ruledtabular}
\begin{tabular}{ccc}
&$\mbox{off-shell charmed }$&$\mbox{off-shell $K^*$ }$\\
\hline
$g_{D_sDK^*}$&$3.36\pm0.43$&$3.17\pm0.41$\\
$g_{D_sD^*K^*}$&$4.62\pm0.60$&$4.47\pm0.57$
\end{tabular}
\end{ruledtabular}
\end{table}
The errors are estimated by  variation of the Borel parameters,
variation of the continuum thresholds, the leptonic decay constants
and uncertainties in the values of the other input parameters. It
should be noted that the main uncertainty comes from the continuum
thresholds and the decay constants.

Table \ref{T34} shows a comparison between our results with the
values predicted by the light-cone sum rules (LCSR) method. The
results of Ref. \cite{Wang} have been rescaled according to the
strong form factor definitions in Eq. (\ref{eq12}). It should be
reminded that the value of $g_{D_sDK^*} (g_{D_sD^*K^*})$ in Table
\ref{T34} is an average of the two coupling constant values
$g^{K^*}_{D_sDK^*}(g^{K^*}_{D_sD^*K^*})$ and
$g^{D}_{D_sDK^*}(g^{D}_{D_sD^*K^*})$ in Table \ref{T33}.
\begin{table}[th]
\caption{Values of the strong coupling constant using the 3PSR
(ours) and LCSR approaches.} \label{T34}
\begin{ruledtabular}
\begin{tabular}{ccc}
$g$ & Ours &LCSR \cite{Wang}
\\ \hline
$g_{D_sDK^*}$ &$3.26\pm0.43$& $3.22\pm0.42$\\
$g_{D_sD^*K^*}$ &$4.54\pm0.59$& $4.04\pm0.53$
\end{tabular}
\end{ruledtabular}
\end{table}

In order to investigate the strong coupling constant values via the
$SU_{f}(3)$ symmetry, the mass of the $s$ quark is ignored in all
calculations. In view of the $SU_{f}(3)$ symmetry, the values of the
parameters $A$ and $B$ for the $g_{D_sDK^*}$ and $g_{D_sD^*K^*}$
strong form factors are given in Table \ref{T35} with
$(\delta,\delta')=(0.70,0.50)~\rm GeV$.
\begin{table}[th]
\caption{Parameters appearing in the fit functions for the
$g_{D_sDK^*}$ and $g_{D_sD^*K^*}$ form factors in $SU_{f}(3)$ symmetry with
$(\delta,\delta')=(0.70,0.50)~\rm GeV$.}\label{T35}
\begin{ruledtabular}
\begin{tabular}{ccc}
$\mbox{Form factor}$&$A$&$B$\\
\hline
$g^{K^*}_{D_sDK^*}(Q^2)$&2.33&1.96\\
$g^{D}_{D_sDK^*}(Q^2)$&2.97&34.34\\
$g^{K^*}_{D_{s}D^*K^*}(Q^2)$&3.29&7.37\\
$g^{D^*}_{D_{s}D^*K^*}(Q^2)$&3.17& 19.15
\end{tabular}
\end{ruledtabular}
\end{table}
In addition, considering the $SU_{f}(3)$ symmetry, we obtain the
values of the coupling constants of the vertices  ${D_sDK^*}$ and
${D_sD^*K^*}$ as shown in Table \ref{T36}.
\begin{table}[th]
\caption{The coupling constants of the vertices ${D_sDK^*}$ and
${D_sD^*K^*}$, in $SU_{f}(3)$ symmetry. }\label{T36}
\begin{ruledtabular}
\begin{tabular}{ccc}
&$\mbox{off-shell charmed }$&$\mbox{ off-shell $K^*$}$\\
\hline
$g_{D_sDK^*}$&$3.31\pm0.43$&$3.47\pm0.45$\\
$g_{D_sD^*K^*}$&$3.91\pm0.51$&$3.67\pm0.48$
\end{tabular}
\end{ruledtabular}
\end{table}

It is possible to compare the coupling constant values of
$g_{D_sDK^*}$ and $g_{D_sD^*K^*}$ with $g_{DD\rho}$ and
$g_{D^*D^*\rho}$ respectively, in the $SU_{f}(3)$ symmetry
consideration.
\begin{table}[th]
\caption{Values of the  coupling constant using the LCSR, and 3PSR.} \label{T37}
\begin{ruledtabular}
\begin{tabular}{cccc}
$g$ & Ours &3PSR \cite{MChiapparini,Rodrigues3,RDMatheus}& LCSR
\cite{Wang}
\\ \hline
$g_{D_sDK^*}$ &$3.39\pm0.44$& $3.42\pm0.44$& $2.62\pm0.66$\\
$g_{D_sD^*K^*}$ &$3.79\pm0.49$&$4.11\pm0.44$ & $3.56\pm0.60$
\end{tabular}
\end{ruledtabular}
\end{table}

An example of specific application of these coupling constants is in
branching ratio calculations of  $B$ meson decays. It is reminded
that re-scattering effects play an important role in the hadronic
$B$ decays. It is not easy  to take them into account in a
systematic way due to the non-perturbative nature of the
multi-particle dynamics. In practical calculations, the
phenomenological models can be used to overcome the difficulty
\cite{Wang}. The one-particle-exchange is one of these
phenomenological models. In this model, the soft re-scattering of
the intermediate states in two-body channels with one-particle
exchange makes the main contributions. The phenomenological
Lagrangian contains many input parameters, which describe the strong
couplings among the charmed mesons in the hadronic $B$ decays.

For instance, we would like to consider the branching ratio of the
$B^+\to {K^*}^0 \pi^+$ decay according to the method of Refs.
\cite{Isola1,Isola2}. It should be noted that our main goal in this
investigation is to illustrate the use of the coupling constants
$g_{D_sDK^*}$ and $g_{D_sD^*K^*}$ in branching ratio calculations of
$B$ decays. Therefore, we do not discuss the methods of calculation,
which are presented here.

According to Refs. \cite{Isola1,Isola2}, the $B \to K^* \pi$ decay
amplitude, $A_{K^* \pi}$ contains the short-distance (SD) and the
long-distance (LD) contributions:
\begin{equation}\label{eq34}
{\mathcal{M}}_{K^*  \pi} = {\mathcal{M}}_{SD} + {\mathcal{M}}_{LD}.
\end{equation}

Using the effective Hamiltonian for non-leptonic $B$ decays
\cite{Fleischer} in the factorization approximation, the value of
the SD amplitude is ${\mathcal{M}}_{SD}=1.52\times 10^{-8}$, which
is evaluated by the following formula \cite{Isola1}:
\begin{eqnarray}\label{eq35}
{\mathcal{M}}_{SD}(B^+ \to {K^*}^0 \pi^+) &=& G_F \sqrt{2}
F_1^{B\to\pi}(m_{K^\ast}^2) f_{K^\ast}  m_{K^\ast}  V_{tb}^\ast
V_{ts}  \left [ a_4-\frac{ a_{10} }{ 2 } \right ]
\left(\varepsilon^*\cdot p_B\right).
\end{eqnarray}

\begin{figure}[th]
\includegraphics[width=8cm,height=3cm]{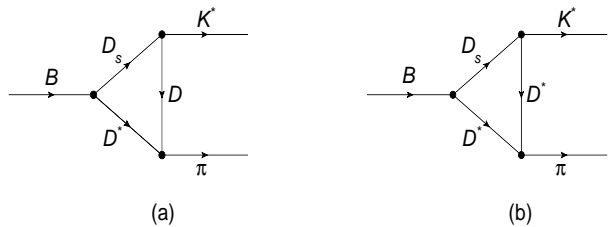}
\caption{Diagrams for the $B\to K^* \pi$ decay with $D_s, D^*$
intermediate states.}\label{F321}
\end{figure}

Fig. \ref{F321} shows diagrams, for the $B\to K^* \pi$ decay with
$D_s, D^*$ intermediate states,  used to calculate the
${\mathcal{M}}_{LD}$ part of the amplitude.
As can be seen in Fig. \ref{F321}, the $B\to {K^*} \pi$ decay may be
occur in two steps. First, the $B$ decays into a $D_s D^*$
intermediate state ($B\to D_s D^*$), and then these two particles
exchange a $D (D^*)$ producing the final $K^*$ and $\pi$. In order
to compute the effect of these interactions in the final decay rate,
we need the $D_s D K^* (D_s D^* K^*)$ and $D^* D \pi (D^* D^* \pi)$
form factors.

The ${\mathcal{M}}_{LD}$ consists of two parts, real and imaginary:
\begin{eqnarray}\label{eq36}
{\mathcal{M}}_{LD}={\mathcal{R}}_{LD}+i\, {\mathcal{I}}_{LD}.
\end{eqnarray}
The computation of the imaginary part of the charming penguin
diagrams contributing to $B\to K^\ast\pi$ decay gives
\begin{eqnarray}\label{eq37}
{\mathcal {I}_{LD}}& =& \frac{m_D}{32 \pi^2 m_B}
\sqrt{\omega^{*2}-1} \int d{\bf n}~ {\cal M}(B\to D_s D^{*}) {\cal
M}(D_s D^{*}\to K^* \pi) ,
\end{eqnarray}
where the integration is over the solid angle. Using  the following
kinematics:
\begin{equation}\label{eq38}
p_B^\mu~=~m_B v^\mu=(m_B,\vec 0 )\ ,\ ~~~~
p^\mu_{D^{*}}~=~m_{D^*}v^{\prime\mu}\ ,\ ~~~~ q~=~p_B-p_{D^{*}}\ ,
\end{equation}
the amplitude for the decay $B\to D_s D^{*}$ is computed by
factorization as:
\begin{eqnarray}\label{eq39}
{\cal M}(B(v)\to D_s(q) D^{*}(\epsilon, v^\prime))&=& - K
\,(m_B+m_{D^*})\,\epsilon^{*}\cdot v\,,
\end{eqnarray}
where $ K =\sqrt{2} \frac{G_{F}}{1+\omega^*}V_{cb}^* V_{cs}\ a_2
\sqrt{m_B m_{D^*}} f_{D_s}$, and $\omega^*  =
\frac{m_B^2+m_{D^*}^2-m^2_{D_s}}{2 m_{D^*} m_B}$. Using the heavy
quark effective lagrangian, the calculation of the amplitude $D_s
\,D^{*}\to K^*\pi$ leads to \cite{Isola1}:
\begin{eqnarray}\label{eq40}
{\cal M}(D_s(q) D^{*}(\epsilon, v^\prime)\to
K^*(p_K,\hat\epsilon)\pi(p_\pi))&=& - \frac{2 g\, F^2(|\vec p_\pi|)
}{f_\pi} \frac{g_V}{\sqrt 2}\sqrt{ \frac{m_{D^*}}{m_{D_s}}}
\epsilon_\lambda\hat\epsilon^*_\sigma \times\left[\frac{2\beta\,
m_D\, q^\sigma p_\pi^\lambda} { (m_{D}v^\prime-p_\pi)^2-m^2_{D}
}\right.\nonumber \\  &+& \left. \frac{4 \lambda\, m_{D^*}\,
G^{\sigma\lambda}(p_\pi,p_K, v^\prime)}{
(m_{D^*}v^\prime-p_\pi)^2-m^2_{D^*} }\right],
\end{eqnarray}
where
\begin{eqnarray}\label{eq41}
G^{\sigma\lambda}(p_\pi,p_K, v^\prime)= &-& (v^\prime\cdot q)
\biggl(g^{\sigma\lambda}(p_K\cdot p_\pi) -p_\pi^\sigma
p_K^\lambda\biggr) - (q\cdot
p_\pi)\biggl(v^{\prime\sigma}p_K^\lambda-g^{\sigma\lambda}(v^\prime\cdot
p_K)\biggr)\nonumber\\ &-& q^\lambda\biggl(p_\pi^\sigma (p_K\cdot
v^\prime) - v^{\prime\sigma} (p_K\cdot p_\pi)\biggr).
\end{eqnarray}
In Eq. (\ref{eq40}),  $F(|\vec p_\pi|)\, = \,0.065$ \cite{Isola2},
$g_V\simeq 5.8$ \cite{Bando}, $g = 0.59 \pm 0.07 \pm 0.01$
\cite{Ahmed}. The basic parameters $\beta$ and $\lambda$ in the
heavy quark effective Lagrangian can be related to the  strong
coupling constants $g_{D_s D K^*}$ and $g_{D_s D^* K^*}$ as
\cite{HQEFT,Wang}:
\begin{eqnarray}\label{eq42}
\beta= \frac{\sqrt{2}\, g_{D_sDK^*}}{2\,g_V},~~~~~~~
\lambda=\frac{\sqrt{2}\, g_{D_sD^*K^*}}{2\,g_V}.
\end{eqnarray}
Our numerical values for the $g_{D_s D K^*}$ and $g_{D_s D^* K^*}$
have been presented in Table \ref{T34}. Using Eqs. (\ref{eq39}) and
(\ref{eq40}) in Eq. (\ref{eq37}) and straightforward calculations,
our numerical value for the imaginary part of the LD amplitude of
two diagrams (a) and (b) in Fig. \ref{F321} is ${\mathcal{I}}^{(a,
b)}_{LD}=-3.81\times 10^{-8}$.

A similar method of the imaginary part is used to calculate the real
part of the LD amplitude \cite{Isola1}. The result for
${\mathcal{R}}^{(a, b)}_{LD}=0.54\times 10^{-8}$, which is the same
order of the imaginary part.

The branching ratio of the non-leptonic process $B^+\to K^{*0}
\pi^+$ is given by
\begin{eqnarray}\label{eq43}
\mathcal{BR}(B^+\to K^{*0} \pi^+) =\frac{\tau_B}{16 \pi
m^3_B}|\mathcal{M}_{K^*\pi}|^2 \sqrt{\lambda(m_B^2, m_K^2,
m_{\pi}^2)},
\end{eqnarray}
where $\lambda (m_B^2, m_K^2,
m_{\pi}^2)=m_B^4+m_K^4+m_{\pi}^4-2m_B^2m_K^2-2m_B^2m_{\pi}^2-2m_K^2m_{\pi}^2$.
Our results for the branching ratio of the  $B^+\to {K^*}^0 \pi^+$
decay are presented in Table \ref{T38}. These results are obtained
for only the short distance amplitude $({\mathcal{M}}_{SD})$, and
also for the total amplitude
(${\mathcal{M}}_{SD}+{\mathcal{M}}_{LD}$). Furthermore, this table
contains the experimental value for the branching ratio of the
$B^+\to {K^*}^0 \pi^+$. Considering the error in the experimental
value, our estimation for the branching ratio value of the $B^+\to
{K^*}^0 \pi^+$ decay with the total amplitude is in consistent
agreement with the experimental data.
\begin{table}[th]
\caption{Branching ratio values (units $10^{-5}$) of the  $B^+\to
{K^*}^0 \pi^+$ mode.} \label{T38}
\begin{ruledtabular}
\begin{tabular}{cccc}
$\rm{}$ & ${\mathcal{M}}_{SD}$ &${\mathcal{M}}_{SD}+{\mathcal{M}}_{LD}$& $\rm{Exp}$\,\cite{Jessop,Eckhart}\\
\hline $\mathcal{BR}(B^+\to {K^*}^0 \pi^+) $ &$0.20\pm 0.03$&
$1.53\pm 0.25$& $1.21 \pm 0.31$
\end{tabular}
\end{ruledtabular}
\end{table}

In summary, taking into account the contributions of the
quark-quark, quark-gluon and gluon-gluon  condensate corrections,
the strong form factors  $g_{D_sDK^*}$ and
 $g_{D_sD^*K^*}$ were estimated within the 3PSR with and without
the $SU_{f}(3)$ symmetry. A comparison was made between  our results and the
predictions of other methods. Finally,  the branching ratio of the
$B^+\to {K^*}^0 \pi^+$ decay was estimated using the coupling
constants of the $D_sDK^*$ and $D_sD^*K^*$ vertices.

\section*{Acknowledgments}
Partial support from the Isfahan University of Technology research
council is appreciated.

\clearpage \clearpage
\appendix
\begin{center}
{\Large \textbf{Appendix--A}}
\end{center}
\setcounter{equation}{0} \renewcommand{\theequation}

In this appendix, the explicit expressions of the coefficients in
the spectral densities are given as:
\begin{eqnarray*}
 I_0(s,s',q^2) &=& \frac{1}{4\lambda^\frac{1}{2}(s,s',q^2)},\nonumber \\
\lambda(a,b,c) &=& a^2+ b^2+ c^2- 2ac- 2bc- 2ac ,\nonumber \\
\Delta&=&s'+m_s^2-m_c^2,\nonumber \\
\Delta'&=&s'+m_c^2-m_s^2,\nonumber \\
\Delta'' &=& s+m_s^2,\nonumber \\
u &=& s+s'-q^2,\nonumber \\
C_1 &=& \frac{1}{\lambda(s,s',q^2)}[2 s' \Delta'' -u \Delta],\nonumber \\
C_2 &=& \frac{1}{\lambda(s,s',q^2)}[2 s \Delta -u \Delta'' ],\nonumber \\
\end{eqnarray*}
also
$C'_1={C_1}_{|_{m_c\leftrightarrow m_s}}$
and $C'_2={C_2}_{|_{m_c\leftrightarrow m_s}}$.

\clearpage
\appendix
\begin{center}
{\Large \textbf{Appendix--B}}
\end{center}
\setcounter{equation}{0} \renewcommand{\theequation}

In this appendix, the explicit expressions of the coefficients of
the quark and gluon condensate contributions of the strong form
factor in the Borel transform scheme is presented.
\begin{eqnarray*}
C_{D_{s}DK^*}^{D}&=&\Bigg(3\,{\frac
{m_{{s}}{m_{{c}}}^{2}}{{M_{{1}}}^{2}}}-3\,{\frac {m_{{s}}{q}
^{2}}{{M_{{1}}}^{2}}}+3\,{\frac
{m_{{c}}{m_{{s}}}^{2}}{{M_{{1}}}^{2}}} -\frac{5}{2}\,{\frac
{{m_{{0}}}^{2}m_{{c}}}{{M_{{1}}}^{2}}}+3\,{\frac {m_{{s}}
{m_{{c}}}^{2}}{{M_{{2}}}^{2}}}-3\,{\frac
{{m_{{0}}}^{2}m_{{c}}}{{M_{{2 }}}^{2}}}-\frac{1}{2}\,{\frac
{{m_{{0}}}^{2}{m_{{c}}}^{3}}{{M_{{1}}}^{2}{M_{{2 }}}^{2}}}\\
&+&\frac{1}{2}\,{\frac
{{m_{{0}}}^{2}m_{{c}}{q}^{2}}{{M_{{1}}}^{2}{M_{{
2}}}^{2}}}+3\,{\frac
{{m_{{c}}}^{3}{m_{{s}}}^{2}}{{M_{{2}}}^{4}}}-\frac{3}{2} \,{\frac
{{m_{{0}}}^{2}{m_{{c}}}^{3}}{{M_{{2}}}^{4}}}\Bigg)\times
e^{-\frac{m_c^2}{M_2^2}},
\end{eqnarray*}
\begin{eqnarray*}
    C_{D_{s}D^* K^*}^{D^*}&=&\Bigg( 6\,{\frac
        {m_{{s}}m_{{c}}}{{M_{{2}}}^{2}}}-6\,{\frac {{q}^{2}{m_{{s}}
            }^{2}}{{M_{{1}}}^{2}{M_{{2}}}^{2}}}+2\,{\frac
        {{m_{{0}}}^{2}{q}^{2}}{{ M_{{1}}}^{2}{M_{{2}}}^{2}}}+6\,{\frac
        {{m_{{c}}}^{2}{m_{{s}}}^{2}}{{M_ {{1}}}^{2}{M_{{2}}}^{2}}}-2\,{\frac
        {{m_{{0}}}^{2}{m_{{c}}}^{2}}{{M_{{ 1}}}^{2}{M_{{2}}}^{2}}}+6\,{\frac
        {{m_{{c}}}^{2}{m_{{s}}}^{2}}{{M_{{2} }}^{4}}}\\&-&3\,{\frac
        {{m_{{0}}}^{2}{m_{{c}}}^{2}}{{M_{{2}}}^{4}}}\Bigg)\times
    e^{-\frac{m_c^2}{M_2^2}},
\end{eqnarray*}
\begin{eqnarray*}
C_{D_sDK^*}^{K^*} &=&-\hat{I}_0(3,2,2)m_c^6-\hat{I}_1(3,2,2)m_c^5 m_s+\hat{I}_2(3,2,2)m_c^5m_s+2\hat{I}_0(3,2,2)m_c^5m_s \\
&&-\hat{I}_1(3,2,2)m_c^4m_s^2+\hat{I}_2(3,2,2)m_c^4m_s^2-2\hat{I}_0(3,2,2)m_c^3m_s^3-\hat{I}_2(3,1,2)m_c^4\\
&&-\hat{I}_0^{[0,1]}(3,2,2)m_c^4+3\hat{I}_1(2,2,2)m_c^4-3\hat{I}_2(2,2,2)m_c^4+\hat{I}_1(3,1,2)m_c^4 \\
&&-3\hat{I}_0(3,2,1)m_c^4+2\hat{I}_2(2,2,2)m_c^3m_s+2\hat{I}_0(3,2,1)m_c^3m_s+2\hat{I}_0^{[1,0]}(3,2,2)m_c^3m_s \\
&&-2\hat{I}_1(2,2,2)m_c^3m_s+\hat{I}_2^{[1,0]}(3,2,2)m_c^3m_s+6\hat{I}_0(4,1,1)m_c^3m_s+4\hat{I}_0(2,2,2)m_c^3m_s \\
&&+\hat{I}_2(3,2,1)m_c^3m_s-\hat{I}_1(3,2,1)m_c^3m_s+\hat{I}_0(3,1,2)m_c^3m_s+\hat{I}_1^{[1,0]}(3,2,2)m_c^3m_s \\
&&-2\hat{I}_0(2,2,2)m_c^2m_s^2+2\hat{I}_1(3,1,2)m_c^2m_s^2-3\hat{I}_0(4,1,1)m_c^2m_s^2-2\hat{I}_2(3,1,2)m_c^2m_s^2 \\
&&-2\hat{I}_0^{[0,1]}(3,2,2)m_c^2m_s^2+5\hat{I}_0(3,1,2)m_cm_s^3+2\hat{I}_2(2,1,3)m_cm_s^3+2\hat{I}_2(3,1,2)m_cm_s^3 \\
&&-2\hat{I}_1(2,1,3)m_cm_s^3-2\hat{I}_1(3,1,2)m_cm_s^3-\hat{I}_0(2,2,2)m_s^4+\hat{I}_0^{[1,0]}(3,2,2)m_s^4 \\
&&-5\hat{I}_1^{[1,1]}(3,2,2)m_c^2+2\hat{I}_0^{[0,1]}(3,1,2)m_c^2-5\hat{I}_2^{[1,1]}(3,2,2)m_c^2+3\hat{I}_2^{[1,0]}(3,2,1)m_c^2 \\
&&+2\hat{I}_0^{[0,1]}(2,2,2)m_c^2+3\hat{I}_1(1,3,1)m_c^2-2\hat{I}_0(1,2,2)m_c^2+3\hat{I}_1^{[1,0]}(3,2,1)m_c^2 \\
&&-3\hat{I}_2(1,3,1)m_c^2-12\hat{I}_0(1,1,3)m_cm_s+\hat{I}_2^{[1,1]}(3,2,2)m_cm_s-3\hat{I}_1^{[1,0]}(3,2,1)m_cm_s \\
&&-9\hat{I}_0(2,1,2)m_cm_s+3\hat{I}_1(1,3,1)m_cm_s-2\hat{I}_1^{[0,1]}(2,1,3)m_cm_s-\hat{I}_2(2,2,1)m_cm_s \\
&&-3\hat{I}_2^{[1,0]}(3,2,1)m_cm_s+\hat{I}_1(2,2,1)m_cm_s+4\hat{I}_1(2,1,2)m_cm_s-2\hat{I}_2^{[0,1]}(2,1,3)m_cm_s \\
&&-4\hat{I}_2(2,1,2)m_cm_s-3\hat{I}_2(1,3,1)m_cm_s+\hat{I}_1^{[1,1]}(3,2,2)m_cm_s-2\hat{I}_0(3,1,1)m_s^2 \\
&&+\hat{I}_2(2,1,2)m_s^2-\hat{I}_1(2,1,2)m_s^2+2\hat{I}_0^{[1,0]}(2,2,2)m_s^2+3\hat{I}_0(1,1,3)m_s^2 \\
&&+\hat{I}_2^{[1,0]}(2,2,2)m_s^2-2\hat{I}_0(2,2,1)m_s^2+\hat{I}_1^{[1,0]}(2,2,2)m_s^2+2\hat{I}_0^{[0,1]}(2,2,2)m_s^2 \\
&&+3\hat{I}_0(2,1,2)m_s^2-2\hat{I}_0^{[1,1]}(2,2,2)-2\hat{I}_0^{[0,2]}(3,1,2)-2\hat{I}_1(1,2,1)+2\hat{I}_0^{[0,1]}(2,2,1) \\
&&+\hat{I}_0^{[0,1]}(1,2,2)+2\hat{I}_2(1,2,1)-3\hat{I}_0^{[1,1]}(3,2,1)-3\hat{I}_0^{[0,1]}(1,1,3)+\hat{I}_0(2,1,1) \\
&&+2\hat{I}_2(1,1,2)+\hat{I}_0^{[0,1]}(2,1,2)-2\hat{I}_1(1,1,2),
\end{eqnarray*}
\begin{eqnarray*}
    C_{D_{s}D^*K^*}^{K^*}&=&2\hat{I}_{2}(3,2,2)m_{c}^{5}+2\hat{I}_{0}(3,2,2)m_{c}^{5}+2\hat{I}_{1}(3,2,2)m_{c}^{5}-2\hat{I}_{2}(3,2,2)m_{c}^{4}m_{s}
    \\
    &&-2\hat{I}_{0}(3,2,2)m_{c}^{3}m_{s}^{2}-2\hat{I}_{2}(3,2,2)m_{c}^{3}m_{s}^{2}-2\hat{I}_{1}(3,2,2)m_{c}^{3}m_{s}^{2}+2\hat{I}_{2}(3,2,2)m_{c}^{2}m_{s}^{3}
    \\
    &&+4\hat{I}_{2}(2,2,2)m_{c}^{3}+2\hat{I}_0^{[0,1]}(3,2,2)m_{c}^{3}+6\hat{I}_{1}(4,1,1)m_{c}^{3}+2\hat{I}_2^{[1,0]}(3,2,2)m_{c}^{3}
    \\
    &&+6\hat{I}_{0}(4,1,1)m_{c}^{3}+4\hat{I}_{0}(2,2,2)m_{c}^{3}+2\hat{I}_{0}(3,2,1)m_{c}^{3}+2\hat{I}_0^{[1,0]}(3,2,2)m_{c}^{3}
    \\
    &&+2\hat{I}_1^{[0,1]}(3,2,2)m_{c}^{3}+2\hat{I}_2^{[0,1]}(3,2,2)m_{c}^{3}-2\hat{I}_{1}(3,1,2)m_{c}^{3}+2\hat{I}_{1}(3,2,1)m_{c}^{3}
    \\
    &&+2\hat{I}_1^{[1,0]}(3,2,2)m_{c}^{3}+4\hat{I}_{1}(2,2,2)m_{c}^{3}+6\hat{I}_{2}(4,1,1)m_{c}^{3}-4\hat{I}_{2}(2,2,2)m_{c}^{2}m_{s}
    \\
    &&-2\hat{I}_2^{[1,0]}(3,2,2)m_{c}^{2}m_{s}+8\hat{I}_{0}(2,1,3)m_{c}^{2}m_{s}+2\hat{I}_{1}(3,1,2)m_{c}^{2}m_{s}-2\hat{I}_{0}(3,1,2)m_{c}^{2}m_{s}
    \\
    &&+4\hat{I}_{2}(2,1,3)m_{c}^{2}m_{s}-2\hat{I}_{2}(3,1,2)m_{c}^{2}m_{s}-6\hat{I}_{2}(4,1,1)m_{c}^{2}m_{s}-2\hat{I}_2^{[0,1]}(3,2,2)m_{c}^{2}m_{s}
    \\
    &&+4\hat{I}_{1}(3,1,2)m_{c}m_{s}^{2}+4\hat{I}_{0}(3,1,2)m_{c}m_{s}^{2}-2\hat{I}_1^{[1,0]}(3,2,2)m_{c}m_{s}^{2}+12\hat{I}_{1}(1,1,4)m_{c}m_{s}^{2}
    \\
    &&-2\hat{I}_2^{[1,0]}(3,2,2)m_{c}m_{s}^{2}+12\hat{I}_{2}(1,1,4)m_{c}m_{s}^{2}-2\hat{I}_0^{[1,0]}(3,2,2)m_{c}m_{s}^{2}+6\hat{I}_{2}(3,1,2)m_{c}m_{s}^{2}
    \\
    &&+12\hat{I}_{0}(1,1,4)m_{c}m_{s}^{2}-4\hat{I}_{2}(3,1,2)m_{s}^{3}+2\hat{I}_2^{[1,0]}(3,2,2)m_{s}^{3}-12\hat{I}_{2}(1,1,4)m_{s}^{3}
    \\
    &&-4\hat{I}_{2}(2,1,3)m_{s}^{3}-2\hat{I}_{2}(2,2,2)m_{s}^{3}+2\hat{I}_{2}(3,1,1)m_{c}-4\hat{I}_2^{[1,0]}(3,2,1)m_{c}
    \\
    &&-4\hat{I}_1^{[0,1]}(3,1,2)m_{c}+4\hat{I}_{0}(1,2,2)m_{c}-2\hat{I}_1^{[1,0]}(3,1,2)m_{c}+4\hat{I}_{2}(1,2,2)m_{c}
    \\
    &&+8\hat{I}_{0}(2,1,2)m_{c}+8\hat{I}_{2}(2,1,2)m_{c}+2\hat{I}_{0}(3,1,1)m_{c}+4\hat{I}_{1}(1,2,2)m_{c}
    \\
    &&-6\hat{I}_1^{[1,0]}(3,2,1)m_{c}+2\hat{I}_{0}(2,2,1)m_{c}-4\hat{I}_0^{[0,1]}(3,1,2)m_{c}-4\hat{I}_2^{[0,1]}(3,1,2)m_{c}
    \\
    &&+2\hat{I}_2^{[1,1]}(3,2,2)m_{c}+2\hat{I}_1^{[1,1]}(3,2,2)m_{c}+2\hat{I}_{1}(2,2,1)m_{c}+6\hat{I}_{1}(2,1,2)m_{c}
    \\
    &&+2\hat{I}_0^{[1,1]}(3,2,2)m_{c}-6\hat{I}_0^{[1,0]}(3,2,1)m_{c}-2\hat{I}_{1}(3,1,1)m_{c}+8\hat{I}_1^{[1,0]}(2,1,3)m_{s}
    \\
    &&+4\hat{I}_2^{[1,0]}(3,2,1)m_{s}+12\hat{I}_{1}(1,1,3)m_{s}-4\hat{I}_{2}(1,2,2)m_{s}+2\hat{I}_2^{[0,1]}(2,2,2)m_{s}
    \\
    &&-2\hat{I}_{2}(2,2,1)m_{s}+4\hat{I}_2^{[0,1]}(3,1,2)m_{s}-2\hat{I}_2^{[1,1]}(3,2,2)m_{s}+4\hat{I}_{1}(2,1,2)m_{s}
    \\
    &&+4\hat{I}_{2}(1,1,3)m_{s}+4\hat{I}_{0}(2,1,2)m_{s}-10\hat{I}_{2}(2,1,2)m_{s}+2\hat{I}_2^{[1,0]}(2,2,2)m_{s}
    \\
    &&+4\hat{I}_2^{[0,1]}(2,1,3)m_{s}+20\hat{I}_{0}(1,1,3)m_{s},
\end{eqnarray*}
where
\begin{eqnarray*}
\hat{I}_{\mu}^{[\alpha,\beta]} (a,b,c)&=&
[M_1^2]^{\alpha}[M_2^2]^{\beta}\frac{d^\alpha}{d(M_1^2)^{\alpha}}
\frac{d^\beta}{d(M_2^2)^{\beta}}[M_1^2]^{\alpha}[M_2^2]^{\beta}\hat
I_{\mu} (a,b,c), \nonumber \\ \hat{I}_k(a,b,c) \!\!\! &=& \!\!\! i
\frac{(-1)^{a+b+c+1}}{16 \pi^2\,\Gamma(a) \Gamma(b) \Gamma(c)}
(M_1^2)^{1-a-b+k} (M_2^2)^{4-a-c-k} \, {U}_0(a+b+c-5,1-c-b),
\nonumber \\ \hat{I}_m(a,b,c) \!\!\! &=& \!\!\! i
\frac{(-1)^{a+b+c+1}}{16 \pi^2\,\Gamma(a) \Gamma(b) \Gamma(c)}
(M_1^2)^{-a-b-1+m} (M_2^2)^{7-a-c-m} \, {U}_0(a+b+c-5,1-c-b),
\nonumber\\ \hat{I}_6(a,b,c) \!\!\! &=& \!\!\! i
\frac{(-1)^{a+b+c+1}}{32 \pi^2\,\Gamma(a) \Gamma(b) \Gamma(c)}
(M_1^2)^{3-a-b} (M_2^2)^{3-a-c} \, {U}_0(a+b+c-6,2-c-b), \nonumber\\
\hat{I}_n(a,b,c) \!\!\! &=& \!\!\! i \frac{(-1)^{a+b+c}}{32
\pi^2\,\Gamma(a) \Gamma(b) \Gamma(c)} (M_1^2)^{-4-a-b+n}
(M_2^2)^{11-a-c-n} \, {U}_0(a+b+c-7,2-c-b),
\end{eqnarray*}
where $k=1, 2$, $m=3, 4, 5$ and $n=7, 8$. We can define the function
$U_0(a,b)$ as:
\begin{eqnarray*}
U_0 (a, b) = \int_0^{\infty} dy (y + M_1^2 + M_2^2)^ay^b \exp
[-\frac{B_{-1}}{y} - B_0 - B_1y ],
\end{eqnarray*}
where
\begin{eqnarray*}
B_{-1} &=& \frac{1}{M_2^2M_1^2}(m_s^2(M_1^2 +M_2^2)^2
-M_2^2M_1^2Q^2), \nonumber\\
B_{0} &=& \frac{1}{M_1^2M_2^2}(m_s^2 + m_c^2)(M_1^2+M_2^2)
,\\
B_{1} &=& \frac{m_c^2}{M_1^2M_2^2}. \nonumber
\end{eqnarray*}

\end{document}